\documentclass[5p,times,twocolumn]{elsarticle}

\usepackage{amssymb}
\usepackage{amsmath}

\usepackage[nodots]{numcompress}

\usepackage{textcomp}

\usepackage{lscape}
\usepackage{subfigure}

\journal{Cryogenics}

\begin{document}

\begin{frontmatter}



\title{A Tiltable Single-Shot Miniature Dilution Refrigerator for Astrophysical Applications}


\author[sjm]{Simon J. Melhuish\corref{cor1}}
\ead{Simon.Melhuish@manchester.ac.uk}

\author[lm]{Lorenzo Martinis}

\author[lp]{Lucio Piccirillo}

\cortext[cor1]{Corresponding author}

\address{Jodrell Bank Centre for Astrophysics,
Alan Turing Building,
School of Physics and Astronomy,
The University of Manchester,
Oxford Road,
Manchester, M13~9PL,
United Kingdom
}

\begin{abstract}
We present a $^{3}$He / $^{4}$He dilution refrigerator designed for cooling 
astronomical mm-wave telescope receivers to around 100\,mK.
Used in combination with a Gifford-McMahon closed-cycle refrigerator,
$^4$He and $^3$He sorption-pumped refrigerators, our cryogen-free system
is capable of achieving 2\,$\mu$W cooling power at 87\,mK.
A receiver attached directly to the telescope optics is required to rotate
with respect to the downward direction. This scenario, of variable tilt,
has proved difficult for typical dilution refrigerators, but our design
has a geometry chosen to allow tilt to 45\textdegree\ and beyond.
\end{abstract}


\end{frontmatter}


\newcommand{\Tmin}{$T_{\rm min}$}
\newcommand{\Vstill}{$V_{\rm still}$}
\newcommand{\Tstill}{$T_{\rm still}$}

\section{Introduction}
\label{sect:introduction}

For several decades astronomers have used cryogenic systems to cool receivers operating in the microwave and mm-wave bands.
Cooling to around 10 -- 20\,K, for example, is necessary for state-of-the-art receiver sensitivity at 10\,mm wavelength in systems
with active amplification.
At shorter wavelengths (mm-wave) bolometric detectors are often employed, where a cryogenically-cooled
absorber is illuminated with the incoming mm-wave signal and its temperature measured to determine the signal power.
Depending upon details of the detection system and local conditions, for example sky emission,
this may typically require a cold stage at around 100\,mK for optimum sensitivity.
Such a temperature may be routinely provided in the lab by $^3$He / $^4$He dilution.
However, when there is a requirement for the receiving system to be mounted upon a telescope
that will follow astronomical targets across the sky this can be problematical for
traditional implementations of the dilution refrigerator (DR) because of their complex gas-handling system and limitations in tilt
angle (although $\pm$ 30\textdegree\ operation is claimed by Batey et al.\cite{Batey 2009}
and the Janis ASTRA dilution refrigerator is specified to operate over an asymmetric tilt range). 
A novel design was developed by Benoit and Pujol \cite{Benoit Pujol 1991}, that dispensed with
the need for a still to draw $^3$He out of the mixing chamber, instead employing a ``vortex pump'' effect.
Our approach, for astrophysical applications, is to use a single-shot diluter.

Single-shot $^3$He fridges are commonly employed in mm/sub-mm astronomical instruments for
several reasons (see for example \cite{DallOglio 1991,Bhatia 2000}). When used in combination with
pre-cooling stages -- either $^4$He or $^3$He/$^4$He fridges -- these systems are
characterized by having no pumping on the $^4$He liquid helium bath,
excellent duty cycle and durations (typical 24 hours operation versus 2--4 hours recycling). They
are also usually very compact with no external gas lines. This last characteristic is 
welcomed when designing systems destined to be operated remotely on turning and rotating telescopes.
The single-shot dilution refrigerator described in this paper is an evolution of the triple-stage
sorption refrigerator described in \cite{Bhatia 2000}, which used a double stage $^3$He/$^4$He fridge to
pre-cool a third $^3$He single-shot fridge capable of achieving a few $\mu$W of cooling power
below 300 mK. 
One of the authors has presented a similar design for continuous circulation using condensation pumping \cite{Teleberg 2008}.
However, the geometry of the heat exchanger (a helix of concentric tubes) do not permit a large tilt range.

\section{Theory}
\label{sect:theory}
In conventional DR systems, single-shot operation is normally
achieved by disabling the $^3$He return line. Under these conditions, the $^3$He
gas leaving the still is pumped away and not returned to the mixing chamber. Removing
the heat input produced by the returning $^3$He allows the mixing chamber to
achieve lower temperatures for a limited time. Very few attempts can be found in the
literature describing systems designed to be operated in single-shot mode \cite{Roach 1999}\cite{Teleberg 2006}.

The cooling power $\dot{Q}$ of a conventional fridge can be calculated 
by using the first law:

\begin{equation}
\dot{Q}=\dot{n}_{3}[H_{3D}-H_{3C}]
\end{equation}

where $\dot{n}_{3}$ is the molar rate at which  $^{3}$He is passing from the concentrated to the dilute phase and $H_{3D}$ and $H_{3C}$
are the enthalpies in the diluted and concentrated phases. We may equivalently express this is terms of entropies:

\begin{equation}
\dot{Q}=\dot{n}_{3}T[S_{3D}-S_{3C}]
\label{eqn:QDotS}
\end{equation}

The enthalpies of the $^{3}$He in the concentrated and diluted phases can be determined by integration of
values for the specific heat.

At low temperatures, where specific heat capacities vary linearly with temperature, eq. \ref{eqn:QDotS} leads
to a simple approximation (see for example \cite{Kent 1993} or \cite{Pobell} eq. 7.30):

\begin{equation}
\dot{Q}=84 \dot{n}_{3} T^{2}  \label{cooling power}
\end{equation}

Radebaugh \cite{Radebaugh 1967} quotes a fit to experimental data to with 1\% at temperatures below 40\,mK, 
but by 100\,mK this can be seen (his fig. 14b) to have increased to approximately 15\%.
Further departure from this model is to be expected at the higher temperatures extant during the initial cool-down of a single-shot DR.

In the case of a continuous-circulation DR these equations assume that the enthalpies / entropies of $^3$He leaving and
re-entering the mixing chamber cancel (although in the case of an inefficient heat exchanger this will not be true).
For the single-shot DR we have no contribution from the enthalpy of returning $^{3}$He. 
However, because the mass of fluid in the mixing chamber
is balanced by that in the still there is a net flow of $^{4}$He into the MC as $^{3}$He is removed.
In our model we therefore track both $^{4}$He and $^{3}$He levels in both phases as time passes.

In a single-shot DR we must optimize the amount of $^3$He left in the concentrated phase after the minimum 
operating temperature has
been reached. This will determine the run-time and therefore the duty-cycle of the fridge.

This means that it is important for us to model the cooling power of the fridge during the cool-down
phase. Therefore we need to evaluate entropies or enthalpies at temperatures above the range of the normal approximation.
Also note that we may start with $^3$He concentrations in both phases close to $50\%$.
The normal approximation of a ``concentrated'' phase that is pure $^3$He over a very dilute phase is not valid.

A knowledge of specific heat capacities for helium mixtures is enough to evaluate the enthalpy or entropy balance,
as we can obtain these values through integration.
Experimental specific heat capacity data, as collected by Radebaugh \cite{Radebaugh 1967} and
Kuerten et al. \cite{Kuerten 1985} have been taken by Chaudhry \cite{Chaudhry 2009}
to make polynomial fits.
We employ these as the input to our model.

Starting with our helium mixture at an initial temperature determined by the $^3$He sorption fridge,
given a set molar flow rate of $^3$He (determined by the pumping speed of the sorption pump
and conditions in the still) out of the mixture we may calculate the rate of $^3$He transport
between phases, and the associated rate of cooling. 
Onto this is added a load power, comprising a power representative of an attached load, plus any contributing heat leaks (see below).
This calculation is performed for each time step,
with the molar quantities in each phase tracking the helium transfer. The mixture temperature is
adjusted at each step according to $\Delta Q$ and the heat capacity of the mixture according to the 
polynomial model.

\begin{figure}
 \centering
 \includegraphics[width=\columnwidth]{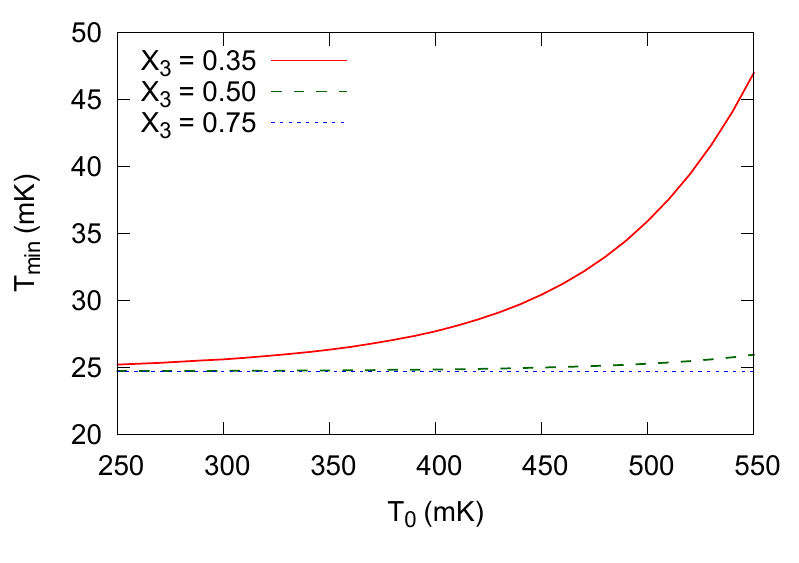}
 \caption{Lowest attainable temperature $T_{min}$ as a function of starting temperature $T_{0}$ for various initial
concentrations of $^3$He and 1\,$\mu$W  load.}
 \label{fig:LowestT}
\end{figure}

\begin{figure}
 \centering
 \includegraphics[width=\columnwidth]{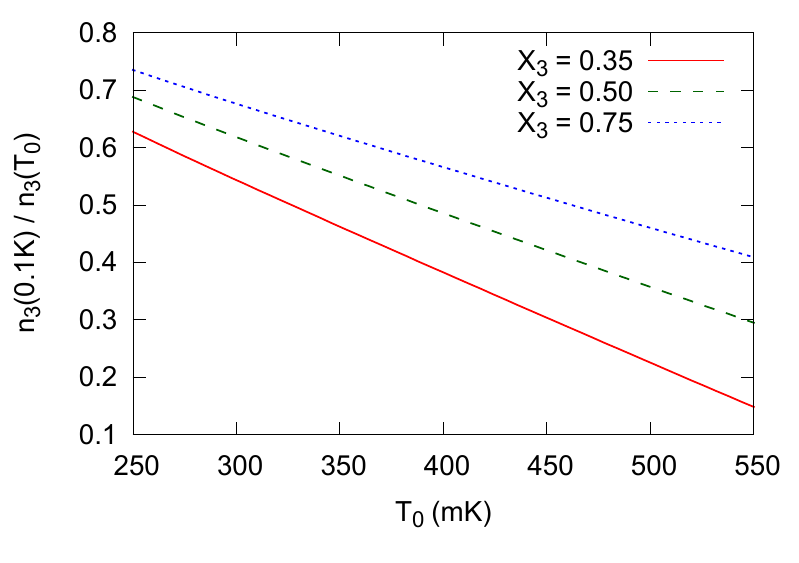}
 \caption{Fraction of initial $^3$He available at 100\,mK after cooling from $T_{0}$ for various initial
concentrations of $^3$He and 1\,$\mu$W load.}
 \label{fig:Remaing3He}
\end{figure}

This results in a minimum-available temperature and the fraction of the initial $^3$He remaining once a running temperature of 100\,mK is reached, 
given the starting temperature.
These are plotted in figures \ref{fig:LowestT} and \ref{fig:Remaing3He}, for three values 
of initial $^3$He concentration.
The balance of extraction rate and load affects the result; for example slower cooling increases the effect of a given
load due to the higher total energy input during the dool-down. A $^3$He transfer rate of 20\,$\mu{\rm mol.s}^{-1}$ has been employed,
with 1 mole of mixture at the start and a simulated total load power of 1\,$\mu$W applied constantly,
which is taken to include all heat leaks.
These plots illustrate that a low initial temperature is
strongly desirable and that the single-shot scheme very much favours higher concentrations
of $^3$He than would be typical for a continuous dilution cooler.

\begin{figure}
 \centering
 \includegraphics[width=\columnwidth]{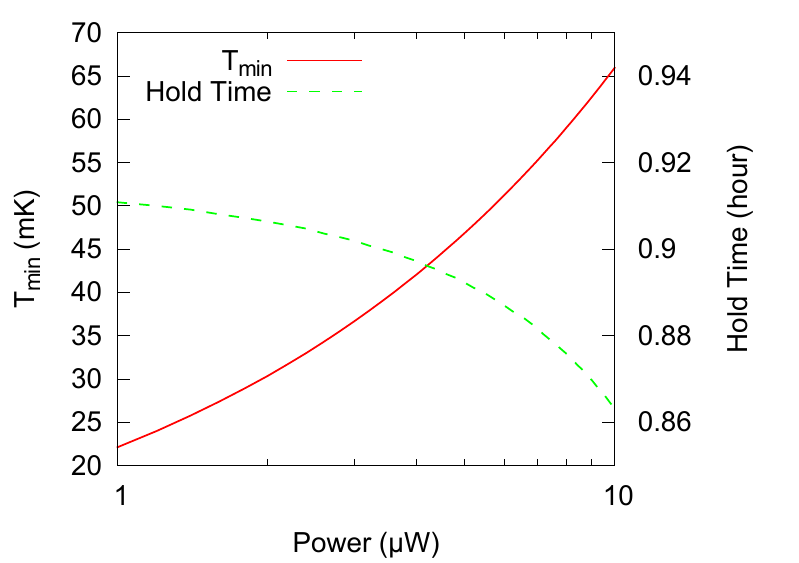}
 \caption{Modelled lowest attainable temperature and hold time as a function of applied power.
The hold times quoted are the time below 100\,mK.}
 \label{fig:ModelledLoadCurve}
\end{figure}

Having established that we need a high $^3$He concentration and a low initial temperature
we examined the effect of applying power to the mixing chamber during the cool-down.
The resulting plot is given in figure \ref{fig:ModelledLoadCurve}.
The $^3$He / $^4$He volumes and $^3$He transfer rate were set to values typical for our system.

\section{Design}
\label{sect:design}

The test installation for the miniature dilution refrigerator (MDR) is built around an RDK-415D Gifford-McMahon (GM) cryo-cooler,
manufactured by Sumitomo Heavy Industries Ltd.
This has specified cooling capacities of 35\,W at 50\,K and 1.5\,W at 4.2\,K
for the first and second stages respectively, when operated from a 50\,Hz supply.
The GM cooler is fixed to a vacuum can machined from aluminium alloy.

A lower-vibration pulse tube cooler would be preferred. However, our PTC was dedicated to another experiment at the time.

\begin{figure}[t]
 \centering
 \includegraphics[width=\columnwidth]{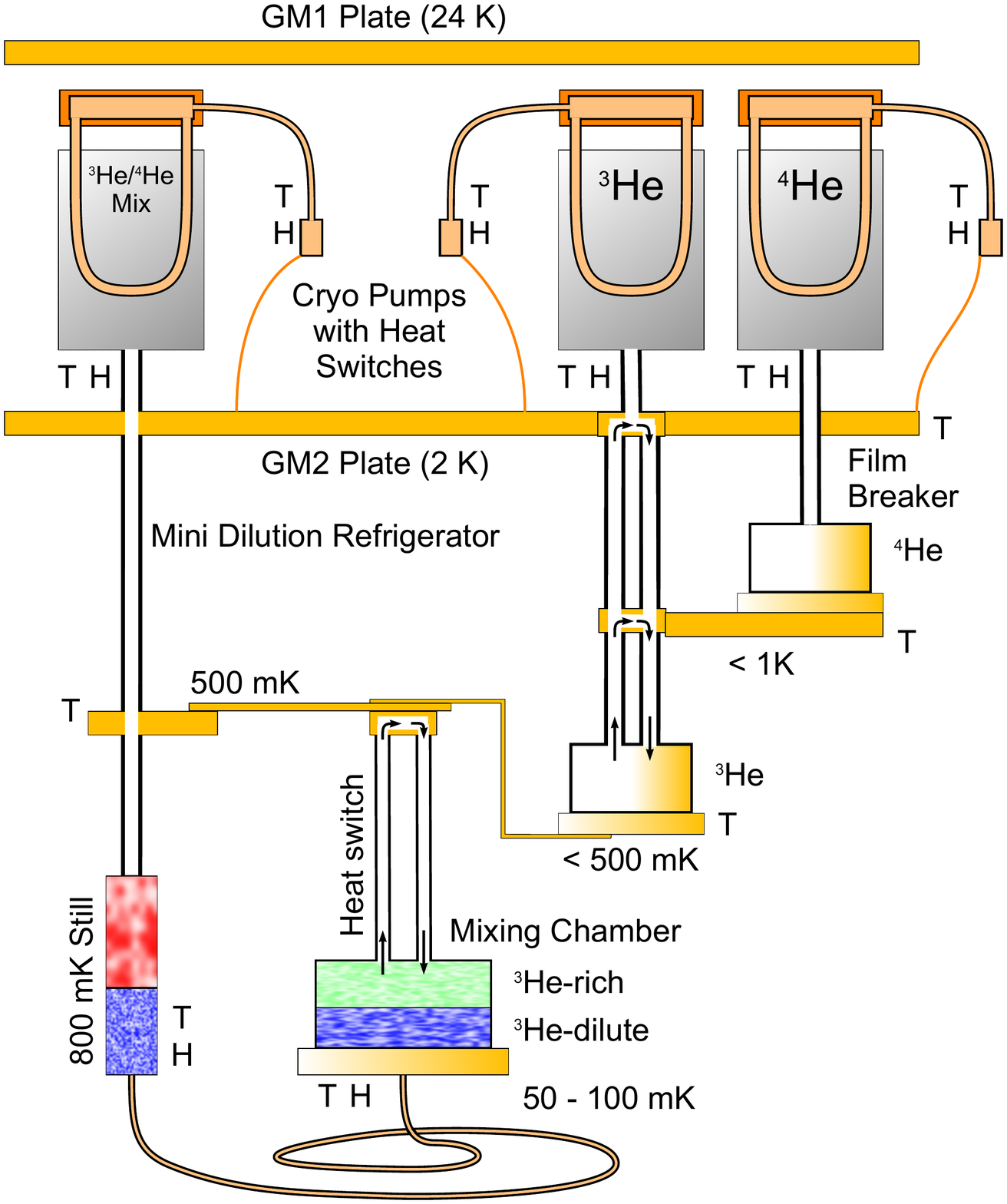}
\caption{Schematic of the MDR system. Thermometers are indicated by a {\it T} and heaters by an {\it H}}
 \label{fig:MDR_system}
\end{figure}

Each GM stage cools a radiation shield fabricated from oxygen-free copper.
The first-stage radiation shield is wrapped within multi-layer insulation --- 20 layers were used.
Typical running temperatures are 23.9\,K and 1.95\,K for the first and second GM stages respectively.
However, these temperatures are elevated by operation of the various cryo pumps.
For example, the first stage temperature may increase to approximately 25\,K, and the second
stage to 5\,K during operation of the large $^{4}$He cryo pump.
A system schematic is presented in figure \ref{fig:MDR_system}.

The second and third stages of cooling are respectively provided by a $^{4}$He and a $^3$He
adsorption cooler.
In each case a charcoal-loaded cryo pump (shaded grey in figure \ref{fig:MDR_system})
is mounted between the first and second GM stages,
and connected by thin-walled stainless-steel tube to an evaporator within the second-stage space.
The cryo pumps are fabricated from 50-mm stainless steel tube with a volume of approximately 200\,ml, filled with 
activated charcoal of approximately 2-mm average grain size.
Heating the cryo pump causes $^4$He / $^3$He to be de-adsorbed; this then condenses in the evaporator.
When the heater is switched off and the cryo pump cooled by means of a heat switch to the GM second-stage
plate, the charcoal starts to re-adsorb helium vapour, thus pumping on, and consequently cooling, the condensed liquid.

To prevent the formation of a $^4$He super-fluid film a {\it film breaker} is inserted at the bottom of
the tube to the $^4$He evaporator.
This consists of an aluminium disc with a pinhole.
An undesirable side-effect of the film breaker is that it prevents free-circulation of helium vapour
during cool-down, which would leave the evaporator too hot to condense helium.
Our approach to solving this problem is to pre-cool the $^4$He  evaporator with a twin-tube linkage
between the $^3$He cryo pump and evaporator.
During pre-cooling of the $^4$He fridge this allows convective circulation of $^3$He gas between 
the GM second stage plate and the $^4$He evaporator connection.
Convective circulation of $^3$He gas from the cold $^4$He evaporator connection
also pre-cools the $^3$He evaporator, allowing efficient condensation of $^3$He.

The final temperatures of the $^4$He and $^3$He evaporators are approximately 0.96\,K and 0.43\,K respectively
for a typical cooling cycle.

The heat switches used to cool the cryo pumps employ vapour convection.
A small cryo pump supplies $^4$He vapour to a twin-pipe circuit between the 
large cryo pump and a copper connection (not shown) to the GM second stage.
This provides sufficient heat transfer to cool the cryo pump by
40\,K (from 55\,K to 15\,K) in approximately 35 minutes.
This switch design will be the subject of a future paper.

The construction of the MDR is similar to that of the adsorption fridges.
Its adsorption cryo pump is located between the first and second GM stages,
with a thin-walled (0.2\,mm wall thickness, 1/4\,inch diameter) tubular stainless steel connection through the second stage cold
plate to the still.
An intermediate cooling connection is made from the tube to the $^3$He evaporator.
Since the still is run at a higher temperature (800\,mK) than the connection to the $^3$He fridge (500\,mK)
the temperature change of helium rising up the tube is inverted.
This is to inhibit any $^4$He super-fluid film, with its associated large conductance.

The still is constructed from copper, providing thermal connections to a ruthenium oxide thermometer and a heater,
supplied with Voltage \Vstill\  from the Dewar controller.
Its internal volume is 8\,ml.
Entering the still at the bottom is its stainless steel capillary connection to the mixing chamber.
The capillary is long and thin (0.8\,mm internal diameter,
0.1\,mm wall thickness, by 200\,mm length), within the limits imposed by consideration of the large
pressures that develop when the system is at ambient temperature, to reduce thermal conduction to the mixing chamber.

The mixing chamber is constructed from copper, with thermal connections for thermometers
(germanium resistance) and a heater for simulating a load.
Its internal volume is 15\,ml.
The capillary connection is to the bottom of the mixing chamber.
The chamber is filled with oxygen-free copper braid to improve the thermal connection
from the chamber walls to the mixture.

At the top of the mixing chamber two further stainless steel tubes 
(110-mm length, 3.175-mm outside diameter and 0.1-mm wall thickness)
are connected to 
a small copper chamber above, in thermal contact with the $^3$He link.
When the MDR is warm and its cryo pump is heated the tubes of the MDR circuit contain $^4$He / $^3$He gas.
Working by convection the twin tubes act as a passive heat switch or \textit{thermal diode}
between the mixing chamber and the $^3$He fridge.
The mixing chamber otherwise has very good thermal isolation and will not cool without this.

Once mixture has been condensed within the mixing chamber,
pre-cooled to the $^3$He fridge temperature by the action of the passive heat switch,
the cryo pump is cooled and the still warmed to start the dilution process.
The mixing chamber then cools further.
When it is colder than the $^3$He fridge convection through the heat switch tubes stops (the switch {\it opens}).

The thermal conduction through the metal tubes of the heat switch plus that through the capillary
to the still is estimated to be 0.1\,$\mu$W. 
Conduction through the residual gas at the saturated vapour pressure of the condensed helium mixture,
is estimated to be below 0.1\,$\mu$W. Other contributing heat leaks are neglected, 
such as conduction through helium between the still and the mixing chamber.

There are no additional radiation shields inside the GM radiation shields, or around the still.

The MDR circuit is charged with 7\,dm$^3$ (STP), or 0.3\,moles, of $^3$He / $^4$He mixture in a 1:1 ratio.
Two 0.4-dm$^3$ expansion volumes are attached to the gas lines for $^3$He and $^3$He / $^4$He mixture
reduce the gas pressure when the system is warm.

\section{Results}
\label{sect:results}

\begin{figure}
 \centering
 \includegraphics[width=\columnwidth]{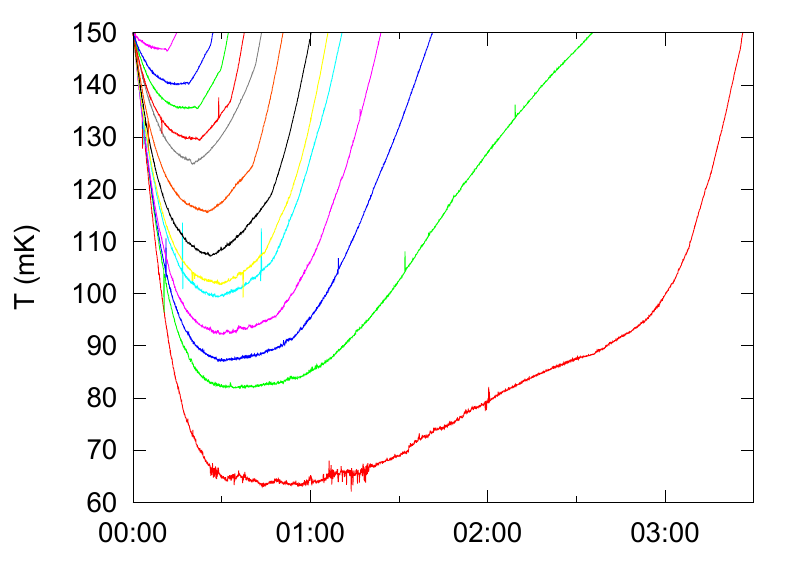}
 \caption{Mixing chamber cooling curves  for applied powers of
0 (lowest), 1, 2, 3, 4, 5, 6, 8, 10, 12, 14, 16 and 18\,$\mu$W (highest).
For no load the mixing chamber can stay below 100\,mK for approximately 3 hours.}
 \label{fig:CoolingCurves}
\end{figure}

Figure \ref{fig:CoolingCurves} shows the temperature of the MDR mixing chamber during several coolings.
The different curves show the results for different levels of power applied to the mixing chamber during
the cooling, to simulate different load conditions.
The cooling that reaches the lowest temperature and stays cold the longest is for zero applied power.
The warmest / shortest cooling had 18\,$\mu$W applied.
Thus we illustrate that we can achieve temperatures below 70\,mK and can hold below 100\,mK for up
to approximately 3 hours.
The total time taken to cycle the sorption fridges and the MDR is approximately 7.5\,hours.
These coolings are with the system oriented level.

Comparison with our model suggests that after the initial cool-down there is
$^3$He remaining in the mixing chamber.
This leads to the flat portion of the cooling curve.
The temperature starts to rise when the $^3$He is exhausted.

Examination of figure \ref{fig:ModelledLoadCurve} suggests that our lowest temperature of approximately
63\,mK implies a heat leak power of 9.2\,$\mu$W applied to the mixing chamber.
This is far higher than our estimates of residual conduction (0.2\,$\mu$W) or radiation.
Whilst we cannot be certain of the actual $^3$He flow rate, and the composition of the mixture following
the pump-down by the cryo-pump, this large discrepancy suggests that there is some other significant source of heating.


Next we examine the effect of still temperature.
For our measurements the fridge was operated at constant \Tstill\  rather
than constant \Vstill.
This has the advantage that as the level of mixture in the still and/or the evaporation rate of the mixture varies, 
there will not be a catastrophic change in still temperature.
Under a constant \Vstill\  it is possible for temperature fluctuations in the mixture, particularly towards the end
of a run when the $^3$He is nearing exhaustion, to cause a run-away increase in \Tstill.

\Tstill\  is held constant by a software PID running on the Dewar controller.
This increases or decreases \Vstill\  accordingly as \Tstill\  falls or rises.

\begin{figure}
 \centering
 \includegraphics[width=\columnwidth]{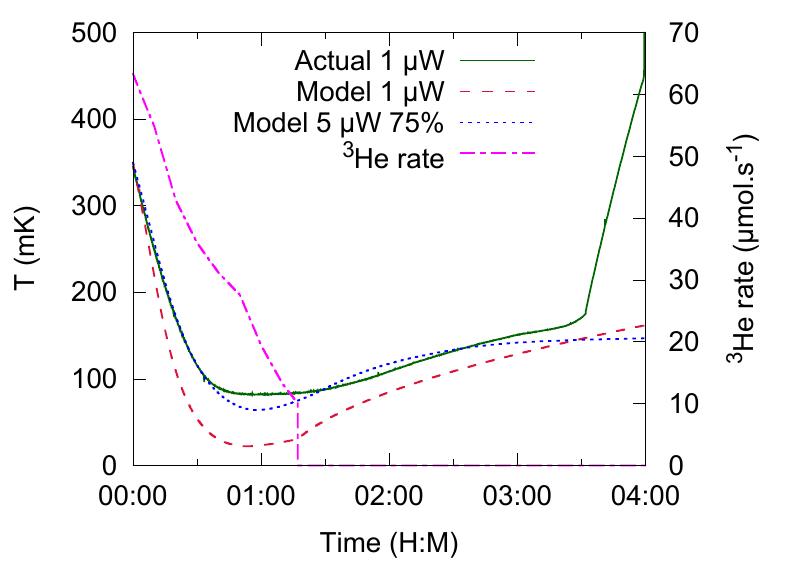}
 \caption{A comparison of an experimental cooling curve with 1\,$\mu$W applied and our modelled results.
The modelled curve with 1\,$\mu$W applied uses the $^3$He transfer rate shown.
A better fit is obtained by reducing the transfer rate to 75\% of this and increasing the load to 5\,$\mu$W.}
 \label{fig:CoolingCurveModel}
\end{figure}

As shown in Radebaugh \cite{Radebaugh 1967} the effective latent heat of evaporation for the helium
in the still varies as a function of both the still and mixing chamber temperatures.
But for reasons of simplicity we have adopted a constant value of 24\,J.mol$^-1$ for our model.
The model input power to the still is taken from real values of an example cooling.
The resulting cooling curve is shown in figure \ref{fig:CoolingCurveModel}, 
along with the corresponding experimental cooling curve.
Also shown is the $^3$He flow rate used in the model.

We see that whilst the model reproduces the general shape of the cooling curve,
including the constant-temperature {\it holding} period, it predicts a faster
initial dool-down and a lower minimum temperature.
To some extent this reflects the uncertainty in initial conditions within the
mixing chamber (MC) --- we do not know how much of the mixture is in the MC
when we start running the still, as some will have evaporated following
operation of the cryo-pump.
However, we find that the model can be made better to agree with the experimental cool-down slope
by rescaling the $^3$He by a factor of $0.75$, and that adding an extra 4\,$\mu$W to the load power
improves agreement between $T_{min}$ values.
As noted above, conduction and radiation are not expected to account for heat leaks of this order.
One unmodelled contribution would be back-flow of $^4$He from the still, through imperfect heat exchange with the out-flowing $^3$He or by frictional heating.
However, we strongly suspect that our passive switch, using mixture from the mixing chamber,
might not be switching ``off'' fully once cold, probably due to film flow.
We intend to replace this with a separate switch for a future version.
Also, we may be suffering from inefficient thermal transfer between the MC walls and 
the mixture, with the MC temperature measured by a thermometer bolted to the outside face. 
No allowance is made for Kapitza resistances in our measurements.
In future we intend to install a sintered heat exchanger to improve this.

\begin{figure}
 \centering
 \includegraphics[width=\columnwidth]{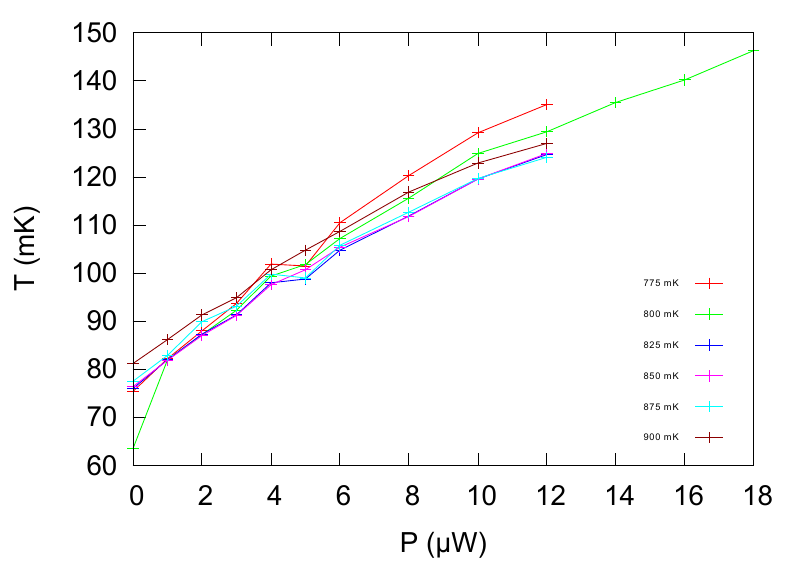}
 \caption{Load curves for values of $T_{\rm still}$ from 775 to 900\,mK up to $P = 12$\,$\mu$W,
and 18\,$mu$W in the case of our adopted $T_{still} = 800$\,mK.}
 \label{fig:LoadCurves}
\end{figure}

\begin{figure}
 \centering
 \includegraphics[width=\columnwidth]{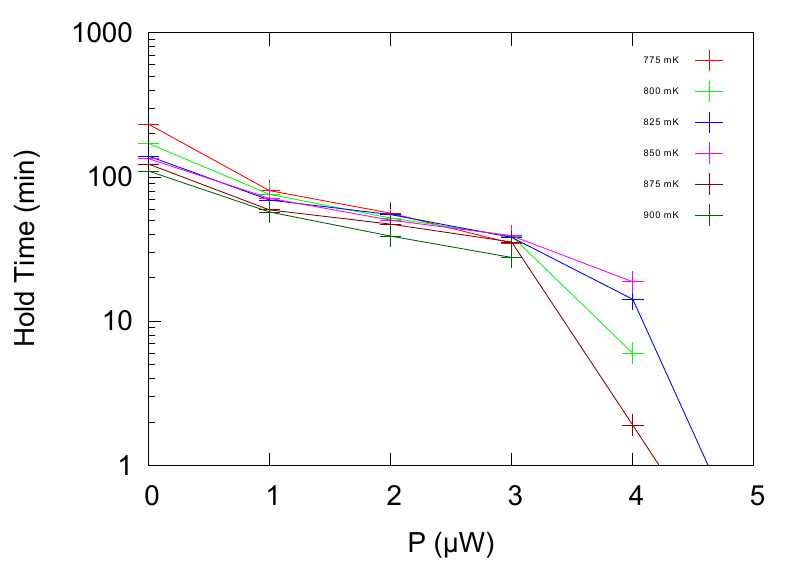}
 \caption{Hold time for values of $T_{\rm still}$ from 775 to 900\,mK}
 \label{fig:HoldTime-vs-TStill}
\end{figure}

Figure \ref{fig:LoadCurves} shows load curves with the system orientation level for
a range of \Tstill\  settings.
Figure \ref{fig:HoldTime-vs-TStill} shows the effect on hold time.
The temperature dependence is not strong.
There is also an effect on the rate of cooling.
A value of 800\,mK was adopted for subsequent experiments.

Our main interest is to be able to maintain a low temperature, with a useful
heat lift, over a range of orientation angles.
Tilting the  MDR to raise or lower the still with respect to the mixing
chamber was found to have a strongly detrimental effect on performance.
Therefore in the experiments that follow only a change in tilt in the perpendicular direction,
i.e. keeping the still and mixing chamber on the same level, was explored.
In each case the  cool-down was performed at the tilt recorded, rather than 
changing the angle once cold.

\begin{figure}
 \centering
 \includegraphics[width=\columnwidth]{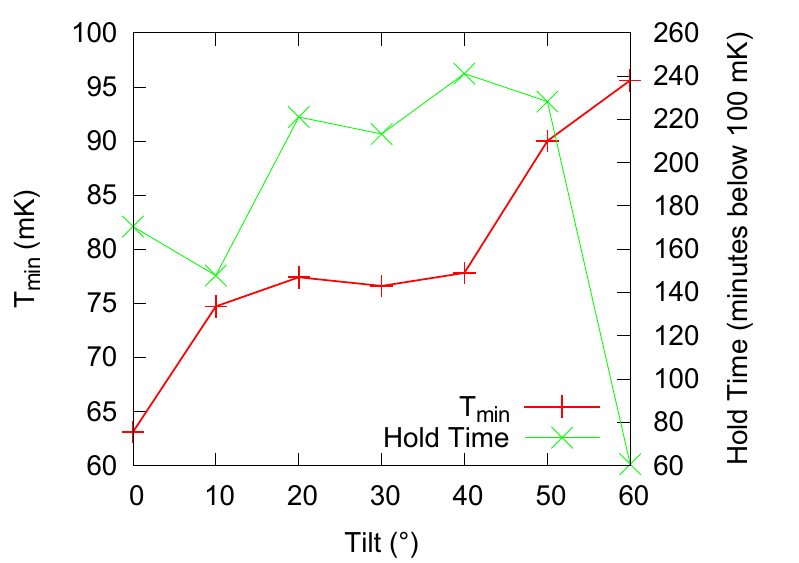}
 \caption{$T_{min}$ and hold time for a range of tilts --- both $T_{min}$ and the hold time
tend to increase with tilt}
 \label{fig:TminLoadTilt}
\end{figure}

Figure \ref{fig:TminLoadTilt} shows the effect of tilt angle on the minimum
temperature and hold time achieved.
For this plot the hold time is measured as the time spent below 100\,mK.
No extra power was applied to simulate a load.
As shown by the solid line there is a general trend of increasing \Tmin\ 
as the fridge is tilted further from level.
However, there is also a general trend of increasing hold time.

\begin{figure}[t]
 \centering
 \includegraphics[width=\columnwidth]{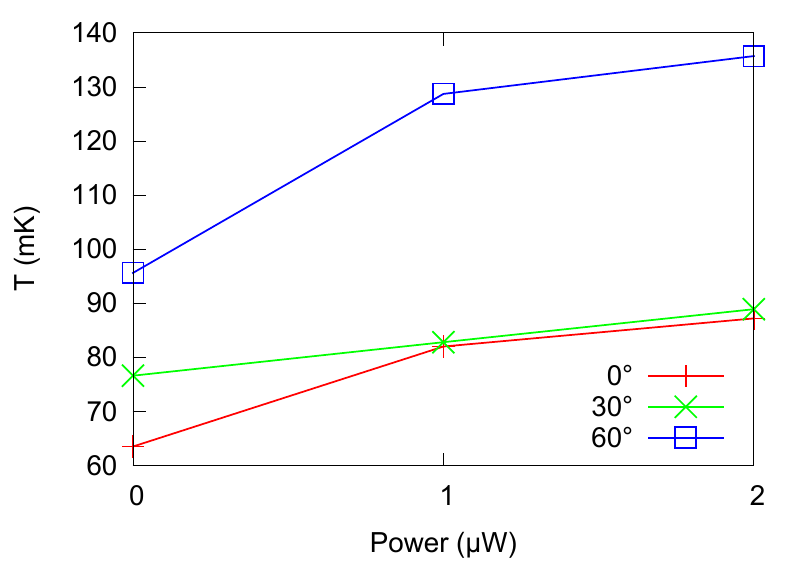}
 \caption{Load curves for 0\textdegree, 30\textdegree\ and 60\textdegree\ tilt}
 \label{fig:TiltLoadCurves}
\end{figure}

\begin{figure}
 \centering
 \includegraphics[width=\columnwidth]{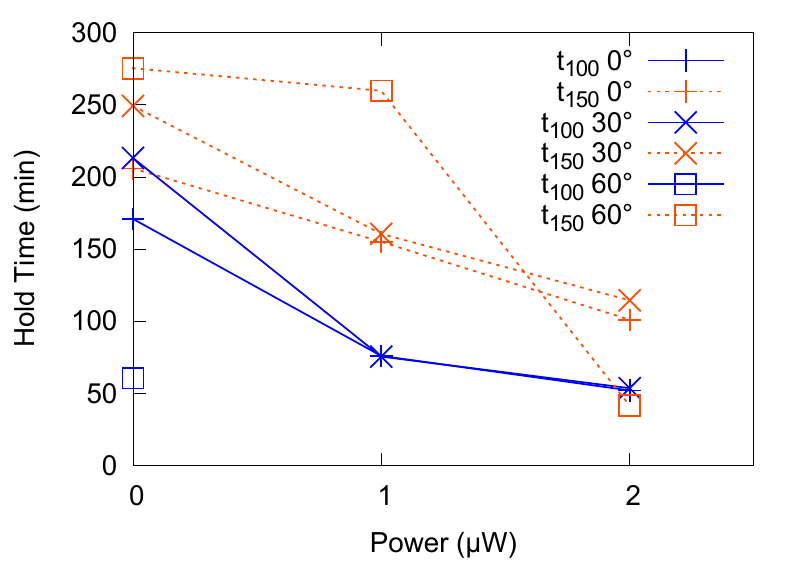}
 \caption{Hold times for 0\textdegree, 30\textdegree\ and 60\textdegree\ tilt ---
Solid lines show hold time below 100\,mK and dashed lines time below 150\,mK.
At 60\textdegree\ tilt only the $P = 0$ run reached below 100\,mK, so $P > 0$
values are not available for 100\,mK hold times}
 \label{fig:TiltHoldTimes}
\end{figure}

To verify that the fridge was still capable of cooling to a useful temperature
under a load condition at these tilts we ran tests at $30^\circ$ and $60^\circ$ tilt
with applied loads of 1\,$\mu$W and 2\,$\mu$W.
Resulting load curves may be seen in Figure \ref{fig:TiltLoadCurves} and
hold times in Figure \ref{fig:TiltHoldTimes}.
As the system is tilted away from level both the running temperature and
hold times increase.

\section{Conclusions}
\label{sect:conclusions}

We have demonstrated a system consisting of a novel $^4$He + $^3$He refrigerator
combined with a prototype single-shot miniature dilution refrigerator,
capable of cooling to 63\,mK with a hold time below 100\,mK of nearly 3 hours.
We have further demonstrated that this system can operate when tilted to $30^\circ$ and
beyond in one direction.

This is itself useful for applications on telescopes, so long as the geometry
allows for the mixing chamber and still to be kept level.
To allow more flexibility of orientation, with a tilt in any direction
about the MDR, we are developing a new concentric design.
This has the still built around the mixing chamber.
It will be the subject of a future paper.
Enlarging the working volumes and using more mixture should also improve performance.
We are hopeful that a significant improvement in cooling power will be realized by
replacing the passive pre-cooling heat switch.



\appendix



\bibliographystyle{model3-num-names}
\bibliography{<your-bib-database>}



\section*{Vitae}

Simon Melhuish graduated in Physics from New College, Oxford, and worked on telescope systems for
Cosmic Microwave Background studies for his PhD at Jodrell Bank, University of Manchester.
He has been responsible for elements of various radio telescope projects, including the 
Very Small Array, QUaD and Clover.

 \includegraphics[width=0.3\columnwidth]{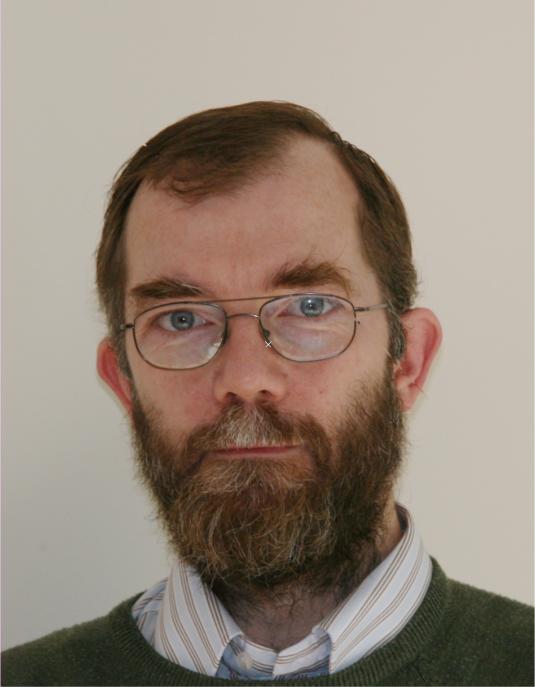}

Lorenzo Martinis studied as a mechanical engineer and worked in the Italian
Nuclear Agency (CNEN) where he was responsible of  hydrogen, helium and
air liquefiers. He has also acquired a remarkable knowledge on projecting and
realizing $^3$He/$^4$He adsorbing and dilution fridges.

 \includegraphics[width=0.3\columnwidth]{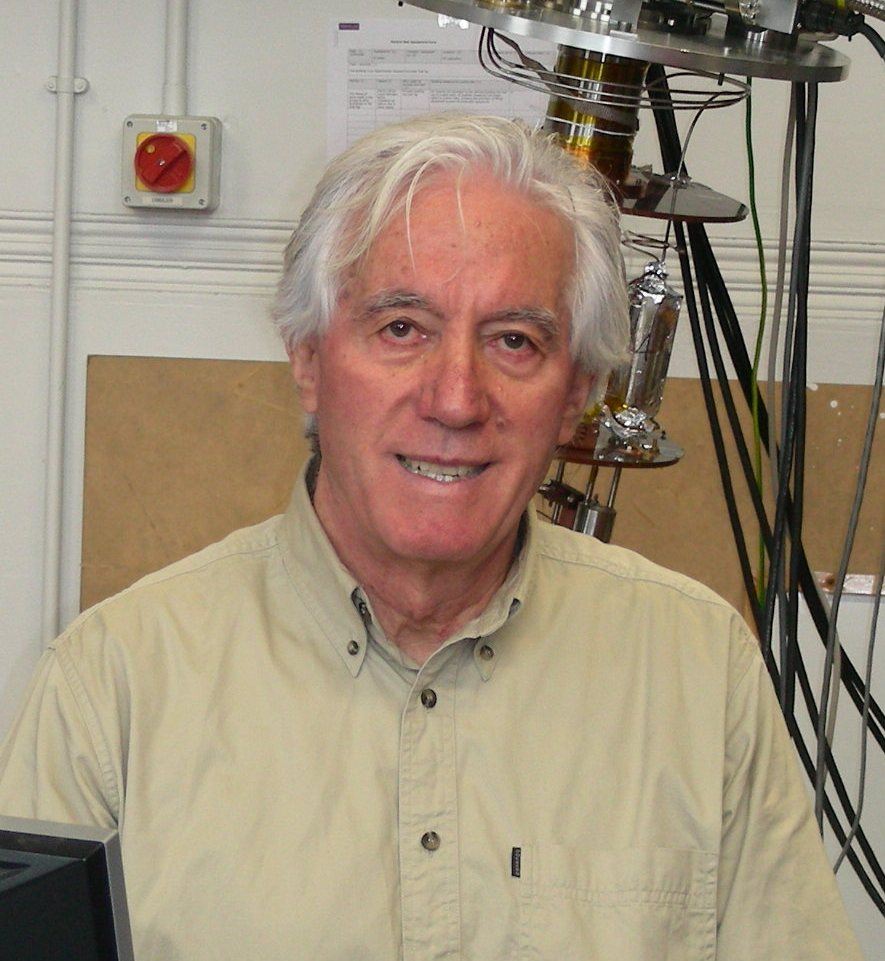}

Lucio Piccirillo is Professor of Radioastronomy Technology at the Jodrell Bank Centre for Astrophysics at the University of Manchester UK.
He is an expert in experimental cosmology, cryogenics, RF devices and novel astrophysical systems. He has more than 130 publications.

 \includegraphics[width=0.3\columnwidth]{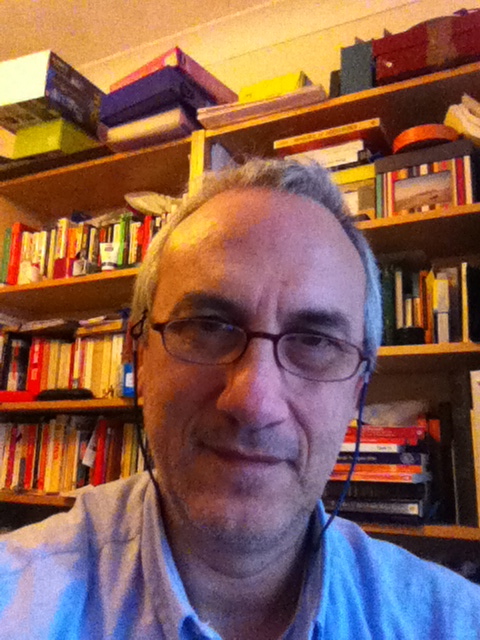}

\end{document}